\documentclass[11pt, a4paper]{article}

\usepackage[utf8]{inputenc}
\usepackage[T1]{fontenc}
\usepackage{amsfonts, amssymb, amsthm, amsmath}
\usepackage{authblk, graphicx, hyperref, enumerate, color}
\usepackage[mathscr]{euscript}
\usepackage{mathrsfs, datetime}
\usepackage[margin=1in]{geometry}

\usepackage{pgfplots}
\pgfplotsset{compat=1.18}
\usepackage{tikz}
\numberwithin{equation}{section}

\newcommand{\beq}{\begin{equation}}
\newcommand{\ene}{\end{equation}}

\theoremstyle{definition}

\newdateformat{daymonthyear}{\THEDAY \ \monthname[\THEMONTH] \ \THEYEAR}

\title{\textbf{ Semiclassical Cosmological Constant in Anti-de Sitter and de Sitter Spacetimes}}

\author[1]{Milton C. Mamani-Leqque\thanks{milton.mamani@unesp.br}}

\affil[1]{\small Instituto de F\'isica Te\'orica, Universidade Estadual Paulista,\par Rua Dr. Bento Teobaldo Ferraz, 271, 01140-070, S\~ao Paulo, Brazil}

\date{\daymonthyear\today}

\begin{document}

\maketitle

% ==============================
% DEDICATORIA
% ==============================

\begin{center}
\emph{A mi esposa Thayss, por su amor y apoyo incondicional.}
\end{center}

% ==============================
% ABSTRACT
% ==============================
 \begin{abstract}
We investigate the semiclassical Einstein equation for a quantized scalar field in four-dimensional anti-de Sitter (AdS) and de Sitter (dS) spacetimes. We first determine self-consistent configurations by fixing the bare cosmological constant to its geometric value, requiring the renormalized vacuum stress-energy tensor to vanish. In AdS, the resulting mass-curvature relations depend on the curvature coupling and boundary conditions; one formal Neumann root for minimal coupling is excluded by the admissible parameter range, whereas both Dirichlet and Neumann solutions are allowed for conformal coupling. In dS, we recover the known self-consistent solution for a minimally coupled massive field. We then evaluate the effective cosmological constant on prescribed backgrounds in four massless configurations. In each case, it takes the form \(\Lambda_{\mathrm{eff}}=A/\ell^2+B/\ell^4\). The competition between classical curvature and vacuum polarization produces a minimum and a zero in the minimally coupled AdS case, while the other configurations are monotonic. Finally, the formal continuation \(\ell\rightarrow i\ell\) fails to reproduce the directly calculated quantum coefficients: their AdS-to-dS ratios are \(29/61\) for minimal coupling and \(-1\) for conformal coupling. These results show that continuation of the curvature scale alone does not establish equivalence between the semiclassical vacuum contributions in AdS and dS.
\end{abstract}

 \section{Introduction}
\label{sec:intro}

The cosmological constant problem remains one of the most profound
challenges at the interface between quantum field theory and
gravitation~\cite{Weinberg1989,Padmanabhan2003,Martin2012}. Within
quantum field theory in curved spacetime, semiclassical gravity
provides a natural framework in which classical spacetime geometry is
coupled to quantum matter through the renormalized expectation value
of the stress--energy tensor~\cite{BirrellDavies,WaldQFT}. It occupies
an intermediate but fundamental position between classical general
relativity and a complete theory of quantum gravity, and offers a
controlled setting in which to investigate how quantum fields modify
gravitational backgrounds. A central difficulty is that the
renormalized stress--energy tensor is not determined solely by the
local curvature. In addition to local geometric information, it can
depend on the quantum state, global properties of the spacetime,
boundary conditions, and infrared structure \cite{WaldQFT,ford2002does}. Understanding the
interplay between these ingredients is therefore essential for
determining which semiclassical geometries can consistently arise from
quantum matter.

Maximally symmetric spacetimes offer a particularly useful arena for
addressing this question \cite{Allen1985,bunch1978quantum}. For suitably symmetric quantum states, their
high degree of symmetry greatly restricts the form of the renormalized
stress--energy tensor and makes the semiclassical Einstein equation
tractable, while still allowing nontrivial dependence on the quantum
field and on the global structure of the background. In four
dimensions, anti-de Sitter (AdS) and de Sitter (dS) spacetimes provide
an especially interesting pair for comparison. Both are maximally
symmetric, characterized by a single curvature scale, and are related
at the level of their classical curvature tensors by the formal
continuation \(\ell\rightarrow i\ell\) \cite{strominger2001ds}. Nevertheless, their quantum
field theories are structurally different. AdS has a timelike
conformal boundary and therefore requires a specification of
admissible boundary conditions, whereas dS is globally hyperbolic and
the Bunch--Davies state provides the standard de Sitter-invariant
choice~\cite{Allen1985}. Consequently, the relation between AdS and
dS at the level of local geometry does not by itself determine a
relation between their quantum stress--energy tensors.

This distinction becomes particularly relevant when one asks what it
means for a maximally symmetric spacetime to be a semiclassical
solution. One possibility is to require the prescribed geometry to
remain an exact solution after quantum effects are included. By fixing
the bare cosmological constant to its classical geometric value
(\(\Lambda_{\rm B} = \Lambda_{\rm geom}\)), this requirement leads to a
self-consistency condition on the parameters of the quantum field and,
for the prescribed geometry considered here, to a vanishing
renormalized stress--energy tensor. Such solutions provide a direct
probe of the extent to which the quantum theory constrains the allowed
field parameters compatible with the prescribed geometry.

A complementary question arises when the quantum contribution is
retained rather than required to vanish. In this case, one may
evaluate the renormalized vacuum energy on a prescribed maximally
symmetric background and determine the corresponding contribution to
the cosmological constant. This perspective is useful for quantifying
vacuum polarization and for studying the relative importance of the
classical curvature and quantum contributions as the curvature scale
is varied. The two viewpoints are conceptually distinct: the first
asks for which field parameters a chosen geometry remains an exact
semiclassical solution, whereas the second asks how the renormalized
quantum stress--energy contributes when the geometry is taken as the
reference background.

These two complementary perspectives provide the starting point for the present analysis, building on previous results. In de Sitter space, the self-consistency problem has been studied by Ju\'arez-Aubry~\cite{JuarezAubry2019}, with a broader classification subsequently developed by Gottschalk \emph{et al.}~\cite{gottschalk2022cosmological}. Conversely, in anti-de Sitter space, Ju\'arez-Aubry and Mamani-Leqque~\cite{JuarezAubryMamani2025} obtained the renormalized vacuum energy density for a scalar field in the Poincar\'e patch, including its dependence on the curvature coupling and on the choice of Dirichlet or Neumann boundary conditions. The present work addresses these complementary sectors by investigating self-consistent solutions in AdS and the effective cosmological contribution in dS, thereby completing the comparison between the two geometries within the framework adopted here.

The present work brings these complementary perspectives together and develops the corresponding analyses in the opposite sectors. We first determine the self-consistent field parameters in AdS, paying particular attention to the role of admissible boundary conditions, and compare them with those obtained in dS. We then retain the renormalized vacuum contribution on the prescribed dS background and study the resulting effective cosmological constant. For the four massless configurations considered here, the results take the common form \(\Lambda_{\rm eff}=A/\ell^2+B/\ell^4\), allowing a direct comparison of the corresponding classical and quantum contributions. Most importantly, we test whether the formal continuation \(\ell\rightarrow i\ell\) reproduces the directly calculated semiclassical contributions. While the classical curvature term transforms as expected, the quantum coefficients do not, in general, match the results obtained independently in AdS and dS. For minimal coupling, the AdS coefficient is \(29/61\) of its dS counterpart; for conformal coupling, the two coefficients have equal magnitude and opposite signs. The aim is not to repeat the underlying renormalization calculations, but to establish what their combined implications reveal about self-consistency, vacuum polarization, and the relation between AdS and dS.

These discrepancies show that the formal relation between the classical curvature scales is insufficient, by itself, to determine the correspondence between the semiclassical vacuum contributions in AdS and dS. The results retain sensitivity to the quantum state, boundary conditions, and global structure of the spacetime. Rather than treating the vacuum contribution merely as a correction to a predetermined cosmological constant, we use the semiclassical Einstein equation itself as the organizing principle and examine the different ways in which quantum matter can be compatible with, or modify, a maximally symmetric geometry. The AdS--dS comparison therefore provides a controlled test of which aspects of a semiclassical solution are controlled by local geometry and which retain information about the underlying quantum theory.

The paper is organized as follows. Section~\ref{sec:eff_cc}
introduces the semiclassical framework and specifies the convention
used for the effective cosmological constant. Section~\ref{sec:ads}
presents the AdS analysis, including the renormalized vacuum energy,
the admissible boundary conditions, the self-consistent solutions,
and the explicit computation of the effective cosmological constant
in the massless minimal and conformal sectors.
Section~\ref{sec:ds} develops the corresponding de Sitter analysis.
Section~\ref{sec:structure} compares the resulting semiclassical
cosmological constants and analyzes their dependence on the curvature
scale. Section~\ref{sec:analytic_cont} examines the relation between
the AdS and dS results under the formal continuation
\(\ell\rightarrow i\ell\). Finally, Section~\ref{sec:conclusions}
summarizes the main results and discusses what the AdS--dS comparison
reveals about the roles of local geometry, quantum state, boundary
conditions, and global structure in semiclassical back-reaction.

% =============================================
\section{Effective Cosmological Constant in Semiclassical Gravity}
\label{sec:eff_cc}
% =============================================

We consider semiclassical gravity on four-dimensional maximally symmetric spacetimes, where the gravitational field is treated classically while quantum matter is described within quantum field theory in curved spacetime. Throughout this work, the backreaction of quantum fields is encoded through the renormalized expectation value of the stress-energy tensor, following the frameworks developed in Refs.~\cite{JuarezAubry2019, JuarezAubryMamani2025}. We use the metric signature $(+,-,-,-)$ throughout.

The dynamics is governed by the semiclassical Einstein equation
\begin{equation}
G_{\mu\nu}+\Lambda_{\rm B}g_{\mu\nu}
=
8\pi G_{\rm N}
\langle T_{\mu\nu}\rangle_{\rm ren},
\label{eq:semi}
\end{equation}
where $\Lambda_{\rm B}$ denotes the bare cosmological constant.

For maximally symmetric quantum states compatible with the spacetime isometries, the renormalized stress-energy tensor is necessarily proportional to the metric,
\begin{equation}
\langle T_{\mu\nu}\rangle_{\rm ren}
=
\rho_{\rm vac}\,g_{\mu\nu},
\label{eq:Tform}
\end{equation}
where $\rho_{\rm vac}$ denotes the renormalized vacuum energy density. 

Inserting \eqref{eq:Tform} into \eqref{eq:semi} gives
\begin{equation}
G_{\mu\nu}
+
\bigl(
\Lambda_{\rm B}
-
8\pi G_{\rm N}\rho_{\rm vac}
\bigr)
g_{\mu\nu}
=
0.
\label{eq:moved}
\end{equation}
This is the standard form in which the renormalized quantum contribution is incorporated into the cosmological term, with
\begin{equation}
\Lambda_{\rm eff}^{\rm (std)}
\equiv
\Lambda_{\rm B}
-
8\pi G_{\rm N}\rho_{\rm vac}.
\end{equation}
In the present work, however, we adopt a convention that keeps the classical curvature contribution and the quantum correction explicit, which is particularly convenient for maximally symmetric spacetimes.

\subsection{Convention used in this work}

For a maximally symmetric spacetime, the Einstein tensor is purely geometric and satisfies
\begin{equation}
G_{\mu\nu}
=
-\Lambda_{\rm geom}\,g_{\mu\nu},
\label{eq:Gclass_gen}
\end{equation}
where $\Lambda_{\rm geom}$ is the classical curvature contribution in $3+1$ dimension:
\begin{equation}
\Lambda_{\rm geom}
=
\begin{cases}
+\dfrac{3}{\ell^{2}}, & \text{de Sitter},\\[0.5em]
-\dfrac{3}{\ell^{2}}, & \text{Anti-de Sitter},
\end{cases}
\label{eq:Lambdageom}
\end{equation}
with $\ell$ the curvature radius. This follows directly from the definition
$G_{\mu\nu}=R_{\mu\nu}-\frac12 R g_{\mu\nu}$
and the fact that
$R_{\mu\nu}=\frac{R}{4}g_{\mu\nu}$
with $R=\pm 12/\ell^2$.

Substituting \eqref{eq:Tform} and \eqref{eq:Gclass_gen} into \eqref{eq:semi} gives
\begin{equation}
-\Lambda_{\rm geom}
+
\Lambda_{\rm B}
=
8\pi G_{\rm N}\rho_{\rm vac},
\end{equation}
hence
\begin{equation}
\Lambda_{\rm B}
=
\Lambda_{\rm geom}
+
8\pi G_{\rm N}\rho_{\rm vac}.
\label{eq:Lambdab_from_Einstein_gen}
\end{equation}

We define the \emph{effective cosmological constant} in this work by
\begin{equation}
\Lambda_{\rm eff}
\equiv
\Lambda_{\rm B}
=
\Lambda_{\rm geom}
+
8\pi G_{\rm N}\rho_{\rm vac}.
\label{eq:LambdaEff_def_gen}
\end{equation}
This convention, also used in Ref.~\cite{JuarezAubryMamani2025}, is adopted to keep the classical curvature contribution and the quantum correction explicit, thereby allowing a direct comparison between the dS and AdS sectors.

Equivalently, we can write
\begin{equation}
\Lambda_{\rm eff}
=
\Lambda_{\rm geom}
+
\Delta\Lambda,
\qquad
\Delta\Lambda
\equiv
8\pi G_{\rm N}\rho_{\rm vac}.
\label{eq:LambdaEff_decomp}
\end{equation}
Thus, in this convention, the quantum correction is represented as an additive shift to the classical curvature contribution $\Lambda_{\rm geom}$.

\footnote{
One may alternatively move the quantum source to the left-hand side and define
\begin{equation}
\Lambda_{\rm eff}^{\rm (std)}
=
\Lambda_{\rm B}
-
8\pi G_{\rm N}\rho_{\rm vac}.
\end{equation}
For a prescribed maximally symmetric background, the two descriptions are related by the semiclassical Einstein equation, which gives
\begin{equation}
\Lambda_{\rm eff}^{\rm (std)}
=
\Lambda_{\rm geom}.
\end{equation}
In the present work we retain the convention in Eq.~\eqref{eq:LambdaEff_def_gen} because it makes the quantum correction explicit as an additive contribution to the classical curvature term and provides a direct comparison with the convention used in Ref.~\cite{JuarezAubryMamani2025}.
}

The specific form of $\rho_{\rm vac}$ (and hence $\Delta\Lambda$) depends on the background spacetime, the quantum state, and the field parameters; these will be derived in the following sections for AdS and dS separately.

For a quantized scalar field, the renormalized vacuum energy density is obtained through Hadamard point-splitting renormalization \cite{decanini2008hadamard},
\begin{equation}
\rho_{\rm vac}
=
\rho_{\rm vac}(m^2,\xi,\ell;\alpha_\mu),
\end{equation}
where $m$ denotes the scalar-field mass, $\xi$ the curvature coupling, $\ell$ the curvature radius of the background spacetime, and $\alpha_\mu$ parametrizes the finite renormalization ambiguity of the stress-energy tensor \cite{hollands2001local, JuarezAubry2019}.\footnote{
See \cite{WaldQFT,BirrellDavies} for general background on Hadamard renormalization in curved spacetime.
}

The parameter $\alpha_\mu$ does not correspond to an observable physical degree of freedom. As shown in Ref.~\cite{JuarezAubryMamani2025}, the finite renormalization ambiguity can be absorbed into suitable redefinitions of the gravitational couplings, namely the bare cosmological constant and Newton's constant, leaving the physical predictions of the semiclassical theory unchanged. After fixing a renormalization prescription, the renormalized vacuum energy becomes uniquely determined by the field parameters, the spacetime geometry, the choice of quantum state, and, when required, boundary conditions.

Although the effective description is common to all maximally symmetric spacetimes, the explicit form of $\rho_{\rm vac}$ depends on the global properties of the background. In Anti-de Sitter spacetime it is determined by the choice of boundary conditions imposed at the conformal boundary, whereas in de Sitter spacetime it depends on the choice of quantum state, which throughout this work will be taken to be the Bunch--Davies vacuum \cite{Allen1985}.

The introduction of the effective cosmological constant therefore provides a unified framework for comparing semiclassical backreaction in Anti-de Sitter and de Sitter spacetimes. The following sections derive the corresponding quantum corrections in each geometry and show that, despite sharing the same effective description, the two spacetimes exhibit qualitatively different semiclassical behaviour.

% =============================================
\section{Anti-de Sitter Spacetime}
\label{sec:ads}
% =============================================

We now specialize the general framework developed in Section~\ref{sec:eff_cc} to four-dimensional Anti-de Sitter spacetime. Unlike de Sitter spacetime, Anti-de Sitter is not globally hyperbolic due to the presence of a timelike conformal boundary. Consequently, the dynamics of quantum fields cannot be formulated from initial data alone and appropriate boundary conditions at conformal infinity are required in order to define a well-posed quantum theory.

The construction adopted throughout this work follows Ref.~\cite{JuarezAubryMamani2025}, which employs the scalar-field two-point function on the Poincaré fundamental domain (PAdS) constructed in Ref.~\cite{dappiaggi2016hadamard} for arbitrary admissible boundary conditions. This spacetime possesses the required hyperbolicity properties once suitable boundary conditions are imposed, allowing the rigorous construction of quasi-free Hadamard states \cite{dappiaggi2016hadamard} and the renormalized stress-energy tensor by means of Hadamard point-splitting.

\subsection{Renormalized Vacuum Energy}

For maximally symmetric Hadamard states, the renormalized stress-energy tensor is proportional to the metric, as stated in Eq.~\eqref{eq:Tform}, so that the complete quantum information is encoded in the vacuum energy density.

For Dirichlet boundary conditions, the Hadamard renormalization procedure yields \footnote{See \cite{pitelli2019boundary} for the renormalized stress-energy tensor in 2D.}
\begin{equation} \label{roADS}
\rho_{\rm vac}^{AdS, (D)}
=
-\frac{1}{2(2\pi)^2}
\left[
w_1^{(D)}
+\frac{3\xi w_0^{(D)}}{\ell^2}
+\frac{m^4}{8}
-
3\left(\xi-\frac16\right)\frac{m^2}{\ell^2}
+
\left(
18\left(\xi-\frac16\right)^2
-\frac1{60}
\right)
\frac1{\ell^4}
\right],
\end{equation}
where \(w_0^{(D)}\) and \(w_1^{(D)}\), which depend on the state, denote the regular Hadamard coefficients computed in Ref.~\cite{JuarezAubryMamani2025}, and have the form:
\begin{align}
w_0^{(D)} &= \frac{2(4\nu^2-1)H_{\nu-\frac12} - 4\nu^2 - \frac13}{8\ell^2}, \label{eq:w0AdS}\\
w_1^{(D)} &= \frac{4(4\nu^2-9)(4\nu^2-1)H_{\nu-\frac12} - 80\nu^4 + \frac{280}{3}\nu^2 + \frac{47}{5}}{256\ell^4}, \label{eq:w1AdS}
\end{align}
with \(\nu \in (0, \infty)\) for Dirichlet. Neumann boundary conditions are obtained through the transformation
\begin{equation}
\nu\rightarrow-\nu,
\qquad
\nu \in (0, 1).
\end{equation}
The dependence on the boundary conditions reflects the absence of a unique vacuum state in Anti-de Sitter spacetime. Although the local ultraviolet structure is universal for all Hadamard states, the infrared behaviour is determined by the choice of boundary conditions at conformal infinity.

\subsection{Spectral Structure}

The Hadamard coefficients depend on the spectral parameter
\begin{equation}
\nu
=
\frac12
\sqrt{
1
+
4m^2\ell^2
-
48\left(\xi-\frac16\right)
}.
\end{equation}
Reality of \(\nu\) is guaranteed by the Breitenlohner--Freedman stability bound \cite{breitenlohner1982stability},
\begin{equation}
m^2\ell^2
\ge
-\frac14
+
12\left(\xi-\frac16\right),
\end{equation}
which ensures that the scalar field lies within the stable sector of the theory. Consequently, the Hadamard coefficients Eqs.~(\ref{eq:w0AdS})--(\ref{eq:w1AdS}) remain real throughout the physically admissible parameter space.

\subsection{Effective Cosmological Constant}

For AdS spacetime, we use the general decomposition \eqref{eq:LambdaEff_decomp} with \(\Lambda_{\rm geom}=-3/\ell^2\):
\begin{equation}
\Lambda_{\rm eff}
=
-\frac3{\ell^2}
+
\Delta\Lambda,
\qquad
\Delta\Lambda
=
8\pi G_{\rm N}\rho_{\rm vac}^{AdS, (D/N)}.
\end{equation} \footnote{Throughout this work, the superscript $D$ denotes Dirichlet boundary conditions and $N$ denotes Neumann boundary conditions.}

\subsection{Self-Consistent Semiclassical Solutions in AdS}
\label{subsec:ads_selfconsistent}

The semiclassical backreaction analysed above determines the quantum-corrected
cosmological constant for a prescribed Anti-de Sitter geometry through
\begin{equation}
\Lambda_{\rm eff}
=
\Lambda_{\rm geom}
+
\Delta\Lambda,
\qquad
\Delta\Lambda
=
8\pi G_{\rm N}\rho_{\rm vac}.
\end{equation}
This analysis allows $\Delta\Lambda$ to be nonzero and describes the general
semiclassical backreaction of the quantum field. Here we consider a
complementary question: whether the quantum field admits parameter values for
which an exact Anti-de Sitter geometry remains a self-consistent solution of
the semiclassical Einstein equations.

This second approach is motivated by the self-consistency analysis developed
for de Sitter spacetime in Ref.~\cite{JuarezAubry2019}. The purpose is to
determine whether the prescribed maximally symmetric AdS geometry can remain
unchanged when the quantum field is included, rather than to assume in
advance that the quantum correction vanishes.

%%%%%%%%%%%%%%%%%%%%%%
For the exact AdS geometry considered here, we now impose the condition
that the prescribed classical geometry remain unchanged after the quantum
field is included. Since the corresponding geometric relations were already
established in Sec.~\ref{sec:eff_cc}, the cosmological term entering the
semiclassical Einstein equation is fixed to the value associated with this
prescribed geometry,
\begin{equation}
\Lambda_{\rm B}
=
\Lambda_{\rm geom}.
\end{equation}

%%%%%%%%%%%%%555%%%%
With this identification, the left-hand side of the semiclassical Einstein
equation vanishes,
\begin{equation}
G_{\mu\nu}
+
\Lambda_{\rm B}g_{\mu\nu}
=
0.
\end{equation}

The semiclassical Einstein equation therefore requires
\begin{equation}
\langle T_{\mu\nu}\rangle_{\rm ren}
=
0.
\label{eq:ads_selfconsistency_condition}
\end{equation}

This condition is not an additional assumption on the quantum correction.
Rather, it follows from requiring the prescribed AdS geometry to remain an
exact solution of the semiclassical Einstein equations. Indeed, combining
$\Lambda_{\rm B}=\Lambda_{\rm geom}$ with the general relation
\eqref{eq:Lambdab_from_Einstein_gen} gives
\begin{equation}
\Lambda_{\rm geom}
=
\Lambda_{\rm geom}
+
8\pi G_{\rm N}\rho_{\rm vac},
\end{equation}
and hence
\begin{equation}
\Delta\Lambda=0,
\qquad
\Lambda_{\rm eff}
=
\Lambda_{\rm geom}.
\end{equation}
Thus, the vanishing of the quantum correction is a consequence of the
self-consistency requirement for the exact AdS geometry, rather than a
condition imposed independently.

This self-consistency analysis does not invalidate the general semiclassical
solutions described by the parameter-dependent
$\Lambda_{\rm eff}=\Lambda_{\rm geom}+\Delta\Lambda$. The two analyses address
different questions. The latter determines the quantum correction to the
cosmological constant for a given field configuration and permits
$\Delta\Lambda\neq0$, whereas the present analysis asks whether there are
distinguished configurations for which the prescribed classical AdS
geometry is preserved exactly. Consequently, both descriptions are valid:
configurations with $\Delta\Lambda\neq0$ are described by the general
backreaction analysis, while configurations satisfying
\eqref{eq:ads_selfconsistency_condition} are the self-consistent solutions
for which the quantum correction vanishes. The latter therefore constitute
the zero-backreaction subset of the general semiclassical description.

To analyse this condition, it is convenient to introduce the dimensionless variable
\begin{equation}
y\equiv m^{2}\ell^{2},
\end{equation}
which combines the field mass and the curvature radius into a single dimensionless parameter.

Using Eqs.~(\ref{eq:w0AdS}) and (\ref{eq:w1AdS}), together with
\begin{equation}
\nu=\sqrt{\frac94+y-12\xi},
\end{equation}
the renormalized vacuum energy assumes the universal form
\begin{equation}
\rho_{\rm vac}^{(D)}
=
-\frac{1}{8\pi^{2}\ell^{4}}
f_{\xi}^{(D)}(y),
\label{eq:rho_ads_dimensionless}
\end{equation}
where
\begin{equation}
\begin{aligned}
f_\xi^{(D)}(y)
&=
\frac{y(4\nu^2-1)}{16}
H_{\nu-\frac12}
\\
&\quad
+\frac1{256}
\left(
-80\nu^4
+\frac{280}{3}\nu^2
+\frac{47}{5}
\right)
-\frac{3\xi}{8}
\left(
4\nu^2+\frac13
\right)
\\
&\quad
+\frac{y^2}{8}
-3\left(\xi-\frac16\right)y
+18\left(\xi-\frac16\right)^2
-\frac1{60}.
\end{aligned}
\label{eq:f_ads_dirichlet}
\end{equation}

After extracting the universal geometric factor \(\ell^{-4}\), all information about the semiclassical vacuum is encoded in the dimensionless function \(f_\xi^{(D)}(y)\). Consequently, the search for exact semiclassical AdS solutions reduces to the algebraic equation
\begin{equation}
f_{\xi}^{(D)}(y)=0.
\label{eq:selfconsistency_ads}
\end{equation}

For a minimally coupled scalar field ($\xi=0$), Eq.~(\ref{eq:selfconsistency_ads}) possesses a unique physically admissible positive solution,
\begin{equation}
y_\ast
=
m^{2}\ell^{2}
=
1.204308835838874\ldots .
\end{equation}
Using \(\Lambda=-3/\ell^2\), this yields the equivalent relation
\begin{equation}
m^{2}
=
-0.4014362786\,\Lambda,
\label{eq:ads_mass_lambda}
\end{equation}
which constitutes the Anti-de Sitter analogue of the de Sitter self-consistency relation obtained in Sec.~\ref{sec:ds}.

The Neumann sector is obtained from the corresponding Dirichlet expressions through the replacement \(\nu\rightarrow-\nu\), leading to the corresponding function \(f_\xi^{(N)}(y)\). For $\xi=0$, the formal solution is

\begin{equation}
y_*=m^{2}\ell^{2}
=
1.84444\ldots.
\label{eq:neumann_formal_root_minimal}
\end{equation}

However, Neumann boundary conditions are admissible only for
\begin{equation}
0<\nu<1,
\end{equation}
which for $\xi=0$ implies
\begin{equation}
-\frac94<y<-\frac54.
\end{equation}
This interval automatically satisfies the Breitenlohner--Freedman bound
\begin{equation}
y\ge -\frac{9}{4},
\end{equation}
and therefore corresponds to stable scalar configurations.

The formal root \eqref{eq:neumann_formal_root_minimal} lies outside the admissible Neumann interval. Consequently, although it satisfies the Breitenlohner--Freedman stability bound, it does not correspond to a physically admissible Hadamard state in the Poincaré fundamental domain.

Therefore, among the admissible Hadamard states considered here for $\xi=0$, only the Dirichlet sector admits an exact self-consistent semiclassical Anti-de Sitter solution. This result illustrates how the timelike conformal boundary of AdS influences the infrared structure of the quantum theory: while Dirichlet and Neumann states share the same local geometry, only the former produces a physically admissible backreaction-free vacuum.

The situation changes significantly for the conformally coupled field ($\xi=1/6$). In this case,
\begin{equation}
\nu=\sqrt{y+\frac14},
\end{equation}
which modifies the admissible parameter space and the structure of the self-consistency condition.

For Dirichlet boundary conditions, Eq.~(\ref{eq:selfconsistency_ads}) again admits a physically admissible solution,
\begin{equation}
y_\ast
=
1.40371873003071\ldots,
\end{equation}
corresponding to
\begin{equation}
m^{2}
=
-0.4679062433\,\Lambda.
\end{equation}

The Neumann sector exhibits qualitatively different behaviour. Solving
\begin{equation}
f_{1/6}^{(N)}(y)=0
\end{equation}
yields the root
\begin{equation}
y_\ast
=
-0.040720880913881\ldots.
\end{equation}
For this value,
\begin{equation}
\nu
=
\sqrt{y_\ast+\frac14}
=
0.45747\ldots,
\end{equation}
which lies within the admissible Neumann interval
\begin{equation}
0<\nu<1.
\end{equation}
Furthermore, the corresponding mass parameter satisfies
\begin{equation}
y_\ast
\ge
-\frac94,
\end{equation}
and therefore respects the Breitenlohner--Freedman stability bound.

Finally the self-consistent solution can be written as
\begin{equation}
m^{2}
=
0.0135736270\,\Lambda.
\end{equation}

Since $\Lambda<0$ in Anti-de Sitter spacetime, the scalar field is slightly tachyonic,
\begin{equation}
m^{2}<0,
\end{equation}
yet remains fully stable owing to the Breitenlohner--Freedman bound. Consequently, unlike the minimally coupled case, the conformally coupled field admits physically admissible self-consistent semiclassical Anti-de Sitter solutions for both Dirichlet and Neumann boundary conditions.

\subsection{Effective Cosmological Constant in the Massless Sector}

We now evaluate the effective cosmological constant in the massless sector. Substituting $m=0$ and the corresponding curvature coupling into the AdS vacuum energy density \eqref{roADS}, and then using the general decomposition \eqref{eq:LambdaEff_decomp} with $\Lambda_{\rm geom}=-3/\ell^2$, one obtains the following results.

For minimal coupling ($m=0$, $\xi=0$), the vacuum energy density reduces to
\begin{equation}
\rho_{\rm vac}^{AdS,(D)}(0,0) = \frac{29}{960\pi^2\ell^4},
\end{equation}
which, upon substitution into \eqref{eq:LambdaEff_decomp}, yields
\begin{equation}
\Lambda_{\rm eff}^{\rm AdS,min}
=
-\frac3{\ell^2}
+
\frac{29G_{\rm N}}
{120\pi\ell^4}.\label{eq:LambdaEff_AdS_min}
\end{equation}

For conformal coupling ($m=0$, $\xi=1/6$), the corresponding vacuum energy density is
\begin{equation}
\rho_{\rm vac}^{AdS,(D)}(0,1/6) = -\frac{1}{960\pi^2\ell^4},
\end{equation}
which gives
\begin{equation}
\Lambda_{\rm eff}^{\rm AdS,conf}
=
-\frac3{\ell^2}
-
\frac{G_{\rm N}}
{120\pi\ell^4}.\label{eq:LambdaEff_AdS_conf}
\end{equation}

These two limiting cases will serve as reference configurations throughout the comparative analysis presented in Section~\ref{sec:structure}.

It is important to note that the Anti-de Sitter massless limits retain the boundary-condition dependence discussed in Ref.~\cite{JuarezAubryMamani2025}. The minimally coupled result corresponds specifically to Dirichlet boundary conditions at the conformal boundary, whereas the conformally coupled result is independent of whether Dirichlet or Neumann boundary conditions are imposed. Accordingly, all references in the following sections to the minimally coupled AdS branch should be understood as referring to the Dirichlet configuration.

 \section{de Sitter Spacetime}
\label{sec:ds}
% =============================================

We now turn to the de Sitter case, following the rigorous construction of the renormalized stress-energy tensor developed in Ref.~\cite{JuarezAubry2019}. Unlike Anti-de Sitter spacetime, de Sitter space is globally hyperbolic and therefore admits a well-posed formulation of quantum field theory without imposing boundary conditions at infinity.

Throughout this work we consider the de Sitter-invariant Bunch--Davies state, which has the Hadamard short-distance structure.
Consequently, once the quantum state and renormalization prescription are fixed, the renormalized expectation value of the stress-energy tensor is determined by the geometric background and the field parameters.

\subsection{Renormalized Vacuum Energy}

For a real scalar field in the Bunch--Davies vacuum, the renormalized vacuum energy density obtained in Ref.~\cite{JuarezAubry2019} is
\begin{equation}
\rho_{\rm vac}^{dS}
=
\frac{1}{(8\pi)^2}
F(m^2,\xi,R)
+
\frac{\alpha_\mu(\xi)}{(8\pi)^2}
m^2\Lambda,
\label{eq:rhovac_ds}
\end{equation}
where the function $F(m^2,\xi,R)$ encodes the quantum fluctuations,
\begin{align}
F(m^2, \xi, R) \equiv & \, m^2\big[m^2 + (\xi - \tfrac{1}{6})R\big] 
                      \left[\psi\!\left(\tfrac{3}{2}+\nu\right) + \psi\!\left(\tfrac{3}{2}-\nu\right) 
                      - \ln\!\left(\frac{12m^2}{R}\right)\right] \nonumber \\
                    & - \frac{1}{2}\left(\xi - \frac{1}{6}\right)^2 R^2 
                      + \frac{R^2}{2160} 
                      - m^2\left(\xi - \frac{1}{6}\right)R 
                      - \frac{m^2 R}{18},
\label{eq:F_ds}
\end{align}
and
\begin{equation}
\nu^2 = \frac{9}{4} - 12\left(\frac{m^2}{R} + \xi\right).
\label{eq:nu_ds}
\end{equation}
As discussed in Sec.~\ref{sec:eff_cc}, the finite renormalization ambiguity $\alpha_\mu$ can be absorbed into a redefinition of the bare gravitational couplings. We therefore set $\alpha_\mu=0$ in what follows.

Note that Eq.~\eqref{eq:rhovac_ds} is consistent with our general convention
\(\rho_{\rm vac}=+F/(8\pi)^2\), with \(F\) defined in Eq.~\eqref{eq:F_ds}.

\subsection{Effective Cosmological Constant}

For de Sitter spacetime, the classical curvature contribution is given by the de Sitter branch of \eqref{eq:Lambdageom}, namely \(\Lambda_{\rm geom}=+3/\ell^2\). Using the general decomposition \eqref{eq:LambdaEff_decomp} together with \eqref{eq:rhovac_ds}, we have
\begin{equation}
\Lambda_{\rm eff}
=
\frac3{\ell^2}
+
\Delta\Lambda,
\qquad
\Delta\Lambda
=
8\pi G_{\rm N}\rho_{\rm vac}^{dS}
=
\frac{G_{\rm N}}{8\pi}F(m^2,\xi,R).
\label{eq:LambdaEff_ds_general}
\end{equation}
This is the de Sitter-specific form of the general relation \eqref{eq:LambdaEff_decomp}.

\subsection{Self-Consistent Semiclassical Solutions}
\label{subsec:ds_selfconsistent}

We now apply the same self-consistency criterion established for the
Anti-de Sitter case: we ask whether the prescribed de Sitter geometry can
remain an exact solution of the semiclassical Einstein equations after the
quantum field is included. Under this requirement, the renormalized quantum
source must vanish,
\begin{equation}
\langle T_{\mu\nu}\rangle_{\rm ren}=0.
\label{eq:ds_selfconsistency_condition}
\end{equation}
Using the renormalized vacuum energy density in Eq.~\eqref{eq:rhovac_ds},
and setting $\alpha_\mu=0$ as above, this condition reduces to
\begin{equation}
F(m^2,\xi,R)=0.
\label{eq:Fzero_ds}
\end{equation}

For a massive minimally coupled scalar field ($\xi=0$), the numerical solution of Eq.~\eqref{eq:Fzero_ds} yields
\begin{equation}
m^2
=
4.89707\times10^{12}\Lambda,
\label{eq:root_ds}
\end{equation}
which constitutes the self-consistent semiclassical de Sitter solution obtained in Ref.~\cite{JuarezAubry2019}. A broader analysis of de Sitter solutions was carried out by Gottschalk \emph{et al.}~\cite{gottschalk2022cosmological}, who showed that the existence and multiplicity of solutions depend sensitively on the curvature coupling and on the renormalization freedom.

\subsection{Effective Cosmological Constant in the Massless Sector}
\label{sec:ds_massless}

The massless limit in de Sitter spacetime requires special care because
the behaviour of the digamma functions in Eq.~\eqref{eq:F_ds} depends
on the curvature coupling. In particular, for minimal coupling one of
the digamma-function arguments approaches a pole as $m\to0^+$,
whereas for conformal coupling both arguments remain finite. We
therefore treat the two cases separately.

\subsubsection{Massless minimally coupled field}
\label{subsec:ds_massless_minimal}

For minimal coupling, $\xi=0$, Eq.~\eqref{eq:nu_ds} becomes
\begin{equation}
\nu
=
\sqrt{\frac94-\frac{12m^2}{R}}
=
\frac32
\sqrt{1-\frac{16m^2}{3R}}.
\label{eq:nu_ds_minimal}
\end{equation}
Hence,
\begin{equation}
\nu\longrightarrow\frac32,
\qquad
\frac32-\nu\longrightarrow0,
\qquad
(m\to0^+).
\end{equation}
Since the digamma function has a simple pole at the origin,
\begin{equation}
\psi(z)
=
-\frac{1}{z}
-\gamma_{\rm E}
+\mathcal{O}(z),
\qquad
z\to0,
\label{eq:psi_pole_ds}
\end{equation}
the limit cannot be obtained by simply setting $m=0$ in
Eq.~\eqref{eq:F_ds}. The leading behaviour of $\nu$ is obtained from
the expansion
\begin{align}
\nu
&=
\frac32
\left(
1-\frac{16m^2}{3R}
\right)^{1/2}
\nonumber\\
&=
\frac32
\left[
1-\frac{8m^2}{3R}
-\frac{32m^4}{9R^2}
+\mathcal{O}(m^6)
\right],
\end{align}
and therefore
\begin{equation}
\frac32-\nu
=
\frac{4m^2}{R}
+\frac{16m^4}{3R^2}
+\mathcal{O}(m^6).
\label{eq:ds_minimal_arg}
\end{equation}
Thus,
\begin{equation}
\psi\left(\frac32-\nu\right)
=
-\frac{R}{4m^2}
+\mathcal{O}(1).
\label{eq:ds_minimal_digamma}
\end{equation}

The singular behaviour in Eq.~\eqref{eq:ds_minimal_digamma} is
precisely compensated by the factor $m^2$ multiplying the digamma
function in Eq.~\eqref{eq:F_ds}. Its finite contribution is
\begin{align}
\lim_{m\to0^+}
m^2\left(m^2-\frac{R}{6}\right)
\psi\left(\frac32-\nu\right)
&=
\lim_{m\to0^+}
m^2\left(m^2-\frac{R}{6}\right)
\left[
-\frac{R}{4m^2}
+\mathcal{O}(1)
\right]
\nonumber\\
&=
\frac{R^2}{24}.
\label{eq:ds_minimal_singular_contribution}
\end{align}
All remaining terms multiplying positive powers of $m^2$ vanish in
the limit. In particular,
\begin{equation}
m^2\ln(m^2)\longrightarrow0,
\qquad
m^2\left(\xi-\frac16\right)R\longrightarrow0,
\qquad
\frac{m^2R}{18}\longrightarrow0.
\end{equation}
The regular contribution from
$\psi(\frac32+\nu)$ also vanishes because
\begin{equation}
\psi\left(\frac32+\nu\right)
\longrightarrow
\psi(3)
=
\frac32-\gamma_{\rm E}.
\end{equation}

Consequently, the strict one-sided limit of the massive expression
in Eq.~\eqref{eq:F_ds} is
\begin{align}
\lim_{m\to0^+}F(m^2,0,R)
&=
\frac{R^2}{24}
-\frac{R^2}{72}
+\frac{R^2}{2160}
\nonumber\\
&=
\frac{61}{2160}R^2.
\label{eq:F_ds_minimal_massless}
\end{align}
For de Sitter space,
\begin{equation}
R=\frac{12}{\ell^2},
\qquad
R^2=\frac{144}{\ell^4},
\end{equation}
and hence
\begin{equation}
\lim_{m\to0^+}F_{\rm dS}(m^2,0)
=
\frac{61}{15\ell^4}.
\label{eq:F_ds_minimal_value}
\end{equation}
Using
$\Delta\Lambda=G_{\rm N}F/(8\pi)$, this gives
\begin{equation}
\Delta\Lambda_{\rm dS,min}^{(m\to0^+)}
=
\frac{61G_{\rm N}}
{120\pi\ell^4},
\label{eq:DeltaLambda_ds_minimal}
\end{equation}
and therefore
\begin{equation}
\Lambda_{\rm eff}^{\rm dS,min}
=
\frac{3}{\ell^2}
+
\frac{61G_{\rm N}}
{120\pi\ell^4}.
\label{eq:LambdaEff_ds_minimal}
\end{equation}

Equation~\eqref{eq:LambdaEff_ds_minimal} is the strict one-sided
limit $m\to0^+$ of the massive Bunch--Davies expression, and should
not be identified with a state-independent statement about the
exactly massless theory. We return to this point in
Sec.~\ref{subsec:ds_massless_conformal}.

\subsubsection{Massless conformally coupled field}
\label{subsec:ds_massless_conformal}

For conformal coupling,
\begin{equation}
\xi=\frac16,
\end{equation}
and Eq.~\eqref{eq:nu_ds} gives
\begin{equation}
\nu^2
=
\frac94
-
12\left(
\frac{m^2}{R}+\frac16
\right)
=
\frac14-\frac{12m^2}{R}.
\label{eq:nu_ds_conformal}
\end{equation}
Therefore,
\begin{equation}
\nu\longrightarrow\frac12,
\qquad
\frac32+\nu\longrightarrow2,
\qquad
\frac32-\nu\longrightarrow1,
\qquad
(m\to0^+).
\end{equation}
Both arguments of the digamma functions remain finite, so no pole is
encountered in this case.

Setting $\xi=1/6$ in Eq.~\eqref{eq:F_ds} gives
\begin{equation}
F\left(m^2,\frac16,R\right)
=
m^4
\left[
\psi\left(\frac32+\nu\right)
+
\psi\left(\frac32-\nu\right)
-
\ln\left(\frac{12m^2}{R}\right)
\right]
+
\frac{R^2}{2160}
-
\frac{m^2R}{18}.
\label{eq:F_ds_conformal}
\end{equation}
Since the digamma functions remain finite and
\begin{equation}
m^4\ln(m^2)\longrightarrow0,
\qquad
m^2\longrightarrow0,
\end{equation}
the first and last terms vanish in the massless limit. Hence,
\begin{equation}
F\left(0,\frac16,R\right)
=
\frac{R^2}{2160}.
\label{eq:F_ds_conformal_massless}
\end{equation}
Using $R=12/\ell^2$,
\begin{equation}
F_{\rm dS}\left(0,\frac16\right)
=
\frac{1}{15\ell^4},
\end{equation}
and therefore
\begin{equation}
\Delta\Lambda_{\rm dS,conf}
=
\frac{G_{\rm N}}
{120\pi\ell^4}.
\label{eq:DeltaLambda_ds_conformal}
\end{equation}
The corresponding effective cosmological constant is
\begin{equation}
\Lambda_{\rm eff}^{\rm dS,conf}
=
\frac{3}{\ell^2}
+
\frac{G_{\rm N}}
{120\pi\ell^4}.
\label{eq:LambdaEff_ds_conformal}
\end{equation}

The two cases therefore differ qualitatively in the way the massless
limit is obtained. For conformal coupling the massive expression has
a regular $m\to0^+$ limit, whereas for minimal coupling the limit is
non-uniform because the argument of one of the digamma functions
approaches its pole.

 It is important to distinguish this limiting procedure from the
construction of the exactly massless, minimally coupled scalar theory
in de Sitter space. Allen~\cite{Allen1985} showed that, for
$m=0$ and $\xi=0$, there is no Fock vacuum invariant under the full
de Sitter group $O(1,4)$. Instead, the infrared sector admits
$E(3)$-invariant states, while the de Sitter-invariant Bunch--Davies
vacuum exists for the massive theory and becomes singular in the
massless limit. Thus, the result obtained above should be understood
specifically as the strict $m\to0^+$ limit of the renormalized
massive Bunch--Davies expression, rather than as an independent
construction of a de Sitter-invariant vacuum for the exactly
massless minimally coupled field.

The appearance of the finite contribution proportional to
$R^2/24$ originates from the pole of
$\psi(3/2-\nu)$ as $\nu\to3/2$, and is therefore absent if one simply
sets $m=0$ before taking the limit. Allen's analysis concerns the
infrared structure of the two-point function and does not determine
the renormalized stress-energy tensor computed here. Consequently,
his result does not by itself fix the numerical coefficient obtained
above, but it makes explicit the infrared subtlety associated with
the massless limit.

The conformally coupled case is free of this infrared singularity,
since $\nu\to1/2$ as $m\to0^+$ and both digamma-function arguments
remain finite. The two massless limits therefore have qualitatively
different infrared behaviour.

These expressions provide the leading semiclassical corrections to the
classical cosmological constant and will be compared with their
Anti-de Sitter counterparts in the following section.

% =============================================
\section{Comparative Structure of the Semiclassical Cosmological Constant}
\label{sec:structure}
% =============================================

The results obtained in Sections~\ref{sec:ads} and \ref{sec:ds} show that, despite their different geometric and global properties, the semiclassical cosmological constants of Anti-de Sitter and de Sitter spacetimes share a common algebraic structure. This observation provides a natural framework for understanding the qualitative features of quantum backreaction and for comparing both geometries on an equal footing.

\subsection{Unified Semiclassical Form}
\label{subsec:unified_form}

For the massless limits discussed in Sections~\ref{sec:ads} and \ref{sec:ds}, the effective cosmological constant can always be written in the form
\begin{equation}
\Lambda_{\rm eff}(\ell)
=
\frac{A}{\ell^{2}}
+
\frac{B}{\ell^{4}},
\label{eq:universal_structure}
\end{equation}
where the coefficient $A$ represents the classical curvature contribution and $B$ encodes the quantum vacuum polarization correction.

The four massless cases considered above correspond to the coefficients listed in Table~\ref{tab:AB}.

\begin{table}[htbp]
\centering
\renewcommand{\arraystretch}{2}
\begin{tabular}{ccc}
\hline
Configuration & $A$ & $B$\\
\hline
dS minimal &
$3$ &
$\dfrac{61\,G_{\rm N}}{120\pi}$
\\[1ex]
dS conformal &
$3$ &
$\dfrac{G_{\rm N}}{120\pi}$
\\[1ex]
AdS minimal &
$-3$ &
$\dfrac{29G_{\rm N}}{120\pi}$
\\[1ex]
AdS conformal &
$-3$ &
$-\dfrac{G_{\rm N}}{120\pi}$
\\
\hline
\end{tabular}
\caption{Coefficients defining the semiclassical cosmological constant through Eq.~(\ref{eq:universal_structure}). The de Sitter minimal coefficient arises from the careful $m\to0^+$ limit of the massive Bunch--Davies vacuum and differs from the conformal case; the conformal de Sitter coefficient is positive, while the corresponding Anti-de Sitter coefficient has opposite sign.}
\label{tab:AB}
\end{table}

Equation~(\ref{eq:universal_structure}) shows that the qualitative behaviour of the massless semiclassical configurations is determined by the relative signs and magnitudes of the coefficients $A$ and $B$. In particular, the competition between the classical contribution $\ell^{-2}$ and the quantum correction $\ell^{-4}$ governs the existence of extrema, sign changes, and the breakdown of the classical approximation.

\subsection{Classification of Backreaction Regimes}
\label{subsec:classification}

The four massless configurations can be classified according to the sign of the product $AB$.

When $AB>0$, the vacuum-polarization correction carries the same sign as the classical curvature contribution. In this case, vacuum polarization reinforces the background geometry and the effective cosmological constant remains monotonic for all values of $\ell$. This behaviour occurs for the minimally coupled de Sitter field, the conformally coupled de Sitter field, and the conformally coupled Anti-de Sitter field.

Conversely, when $AB<0$, the quantum correction opposes the classical curvature. The resulting competition generates nontrivial structure in the semiclassical solutions, including extrema and sign changes. Among the four configurations considered, this occurs only for the minimally coupled Anti-de Sitter field, for which the positive quantum correction counteracts the negative classical contribution.

Thus, the qualitative behaviour of the semiclassical cosmological constant is determined not solely by the sign of the background curvature, but by the relative sign between the classical and quantum contributions. In the cases where $AB>0$, the quantum correction acts as a renormalization of the curvature scale without altering the monotonic character of $\Lambda_{\rm eff}$. In the $AB<0$ case, however, the opposing terms generate a nontrivial structure in $\Lambda_{\rm eff}$, including an extremum and, when the coefficients permit it, a zero of the effective cosmological constant. In the minimally coupled AdS configuration studied here, this structure consists of a global minimum together with a zero at $\ell_0$, as will be analysed
in Secs.~\ref{subsec:competition} and \ref{subsec:critical_scales}.

\subsection{Competition between Classical Curvature and Vacuum Polarization}
\label{subsec:competition}

We now examine explicitly how the competition between classical and
quantum contributions manifests itself in the effective cosmological
constant. In the massless limit, the semiclassical corrections derived
in Sections~\ref{sec:ads} and \ref{sec:ds} reduce to the simple
analytic expressions encoded in Table~\ref{tab:AB}. The classical
contribution scales as $\ell^{-2}$, whereas the quantum correction
scales as $\ell^{-4}$. Consequently, the quantum contribution grows
relative to the classical term as the curvature radius decreases and
eventually becomes dominant at sufficiently small $\ell$.

\begin{figure}[htbp]
    \centering
    \includegraphics[width=0.85\textwidth]{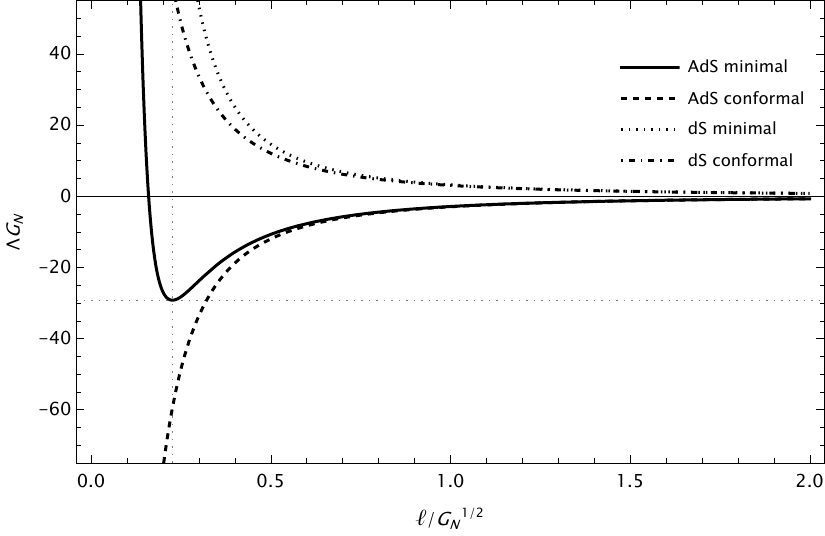}
    \caption{Effective cosmological constant as a function of the
    curvature radius $\ell$ for massless scalar fields in Anti-de
    Sitter and de Sitter spacetimes. Solid and dashed lines correspond
    to minimally coupled ($\xi=0$) and conformally coupled
    ($\xi=1/6$) scalar fields in AdS, respectively, while dotted and
    dash-dotted lines denote the corresponding dS configurations.
    The vertical grid line marks the minimum of the AdS minimal
    branch.}
    \label{fig:Lambda_eff_comparison}
\end{figure}

Figure~\ref{fig:Lambda_eff_comparison} confirms the classification
introduced in Sec.~\ref{subsec:classification}. The four massless
configurations naturally separate according to the sign of the
product $AB$.

As previously noted in Sec.~\ref{subsec:classification}, the three
configurations with $AB>0$ have explicit expressions:

\begin{itemize}
\item \textbf{Minimally coupled de Sitter:}
$\Lambda_{\rm eff}(\ell)=3/\ell^2+61G_N/(120\pi\ell^4)$.
Both terms are positive, so $\Lambda_{\rm eff}>0$ for all $\ell$ and
decreases monotonically with $\ell$.

\item \textbf{Conformally coupled de Sitter:}
$\Lambda_{\rm eff}(\ell)=3/\ell^2+G_N/(120\pi\ell^4)$.
Again, both terms are positive, so $\Lambda_{\rm eff}>0$ for all
$\ell$ and decreases monotonically with $\ell$.

\item \textbf{Conformally coupled Anti-de Sitter:}
$\Lambda_{\rm eff}(\ell)=-3/\ell^2-G_N/(120\pi\ell^4)$.
Both terms are negative, so $\Lambda_{\rm eff}<0$ for all $\ell$ and
increases monotonically, becoming less negative as $\ell$ grows.
\end{itemize}

In all three cases, $\Lambda_{\rm eff}$ remains monotonic, in
agreement with the classification established earlier.

We now consider the remaining configuration, minimally coupled AdS,
which has $AB<0$:
\begin{equation}
\Lambda_{\rm eff}(\ell)
=
-\frac{3}{\ell^2}
+
\frac{29G_N}{120\pi\ell^4}.
\label{eq:Lambda_ads_minimal_explicit}
\end{equation}
Here, the positive vacuum contribution
\[
\Delta\Lambda
=
\frac{29G_N}{120\pi\ell^4}
\]
opposes the negative classical AdS contribution
$-3/\ell^2$. The resulting competition generates a global minimum at
\begin{equation}
\ell_{\min}
=
\sqrt{\frac{29G_N}{180\pi}},
\label{eq:ell_min_ads}
\end{equation}
at which
\begin{equation}
\Lambda_{\rm eff}(\ell_{\min})
=
-\frac{270\pi}{29G_N}
\approx
-29.24\,G_N^{-1}.
\end{equation}
This minimum is clearly visible as the pronounced dip in the black
solid curve of Fig.~\ref{fig:Lambda_eff_comparison}. For
$\ell<\ell_{\min}$, the quantum contribution grows more rapidly than
the classical term as $\ell$ decreases, and $\Lambda_{\rm eff}$
increases as $\ell$ is reduced. The function crosses zero at
\begin{equation}
\ell_0
=
\sqrt{\frac{29G_N}{360\pi}}
=
\frac{\ell_{\min}}{\sqrt{2}}.
\label{eq:ell_zero_ads}
\end{equation}

For $\ell<\ell_0$, $\Lambda_{\rm eff}>0$, while for
$\ell>\ell_0$, $\Lambda_{\rm eff}<0$. Thus, within the effective
description, the two regions are associated with positive and
negative effective cosmological constants, respectively. The sign
change is a direct consequence of the competition between the
classical and quantum contributions.

To better understand the internal structure of this result, it is
instructive to disentangle the purely geometric classical substrate
from the quantum vacuum contribution.

\begin{figure}[htbp]
    \centering
    \includegraphics[width=\textwidth]{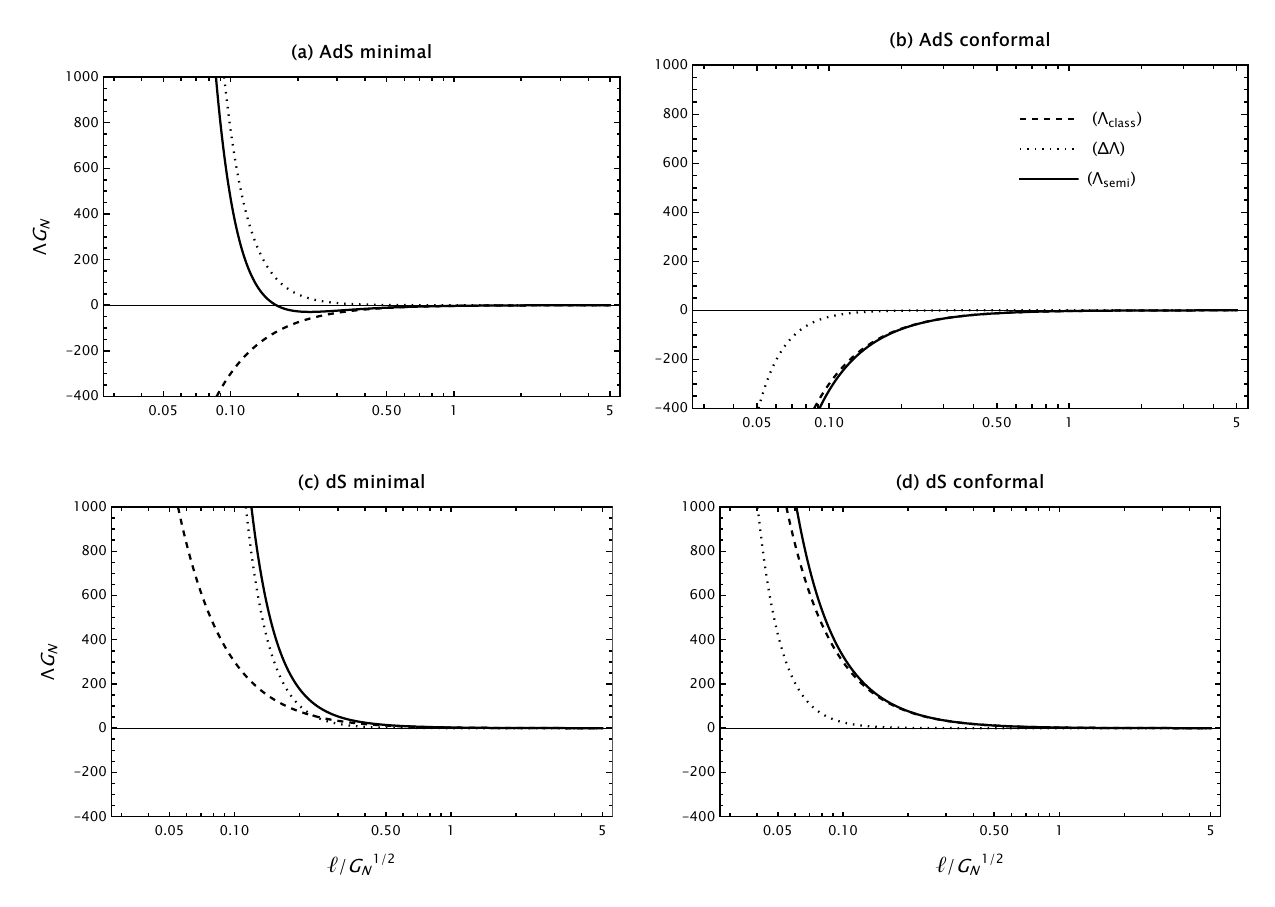}
    \caption{Decomposition of the effective cosmological constant into
    its classical contribution \(\Lambda_{\text{class}}\) (dashed) and
    the quantum vacuum correction \(\Delta\Lambda\) (dotted) for the
    four configurations. The solid line represents the total effective
    cosmological constant,
    \(\Lambda_{\rm eff}=\Lambda_{\text{class}}+\Delta\Lambda\).
    Panels (a) and (b) correspond to AdS minimal and conformal,
    respectively; panels (c) and (d) correspond to dS minimal and
    conformal. Note the relative signs and magnitudes: in panel (a),
    the positive quantum correction competes with the negative classical
    contribution and eventually reverses the sign of their sum at
    sufficiently small \(\ell\). In panels (b), (c), and (d), the
    classical and quantum contributions share the same sign and
    reinforce each other.}
    \label{fig:Lambda_decomposition}
\end{figure} 

The decomposition presented in Fig.~\ref{fig:Lambda_decomposition}
provides a direct visualization of Eq.~\eqref{eq:universal_structure}.
Because the classical term scales as $\mathcal{O}(\ell^{-2})$ and the
quantum correction scales as $\mathcal{O}(\ell^{-4})$, their relative
importance changes systematically with the curvature radius. At
large curvature radii, corresponding to weakly curved macroscopic
spacetimes, the $\ell^{-4}$ term is strongly suppressed, and the total
effective cosmological constant is correspondingly close to the
classical contribution. As $\ell$ decreases, the quantum correction
grows relative to the classical term and eventually becomes dominant
whenever the two contributions have opposite signs and sufficient
magnitude.

A careful examination of the panels reveals that the vacuum
contribution can reverse the sign of the effective cosmological
constant.

In panels (b), (c), and (d) (AdS conformal, dS minimal, and dS
conformal), the quantum correction shares the same sign as the
classical contribution. Here, vacuum fluctuations reinforce the
classical contribution, and $\Lambda_{\rm eff}(\ell)$ remains
monotonic.

In marked contrast, panel (a) (AdS minimal) reveals a direct
competition between the classical geometry and the quantum vacuum.
The positive vacuum contribution
\[
\Delta\Lambda
=
\frac{29G_N}{120\pi\ell^4}
\]
overcomes the negative classical AdS contribution at sufficiently
small $\ell$, leading to a positive value of the effective
cosmological constant,
$\Lambda_{\rm eff}>0$. As $\ell$ increases, the quantum correction
becomes progressively less important, $\Lambda_{\rm eff}$ crosses
zero at $\ell_0$, reaches a negative minimum at $\ell_{\min}$, and
eventually approaches zero from the negative side. In the asymptotic
regime,
\begin{equation}
\Lambda_{\rm eff}\to0^{-},
\end{equation}
so that the effective cosmological constant approaches zero from the
AdS side as $\ell\to\infty$.

These sequences should not be interpreted as dynamical cosmological
histories. Rather, they provide a compact characterization of how the
effective cosmological constant changes as the relative importance of
the vacuum polarization term varies across different regions of
parameter space. Within the effective description adopted here,
positive, vanishing, and negative values of $\Lambda_{\rm eff}$ define
regions with positive, zero, and negative effective cosmological
constant, respectively. If $\Lambda_{\rm eff}$ is instead interpreted
as the cosmological constant entering a fully self-consistent
semiclassical solution, these signs would correspond to de
Sitter-like, flat, and Anti-de Sitter-like branches. However, such an
interpretation requires solving the semiclassical Einstein equation
self-consistently with the corresponding geometry; it does not follow
from the prescribed-background calculation alone.

Nevertheless, the result illustrates that vacuum polarization can
change the sign of the effective cosmological constant and thereby
connect regions of parameter space associated with opposite effective
curvature sectors.

\subsection{Emergence of Critical Scales}
\label{subsec:critical_scales}

The universal form in Eq.~\eqref{eq:universal_structure} allows us to
characterize the structure of $\Lambda_{\rm eff}(\ell)$ by identifying
two distinguished scales: the location of its extrema and its
intersection with the horizontal axis. The first is obtained by
examining the stationary points of the function, while the second is
determined by the values of $\ell$ for which $\Lambda_{\rm eff}$
vanishes.

Differentiating with respect to the curvature radius gives
\begin{equation}
\frac{d\Lambda_{\rm eff}}{d\ell}
=
-\frac{2A}{\ell^{3}}
-\frac{4B}{\ell^{5}}.
\end{equation}
The extrema are therefore determined by
\begin{equation}
\frac{d\Lambda_{\rm eff}}{d\ell}=0,
\end{equation}
which gives
\begin{equation}
\ell_{\rm crit}^{2}
=
-\frac{2B}{A}.
\label{eq:lcrit}
\end{equation}
A real critical scale exists only when $AB<0$. Thus, extrema can
occur only when the classical and quantum contributions have opposite
signs.

Among the four configurations considered, only the minimally coupled
Anti-de Sitter field satisfies this condition. This is the
configuration already analysed in Sec.~\ref{subsec:competition}, for
which $A=-3$ and $B=29G_N/(120\pi)$. In this case, Eq.~\eqref{eq:lcrit}
yields
\begin{equation}
\ell_{\min}
=
\sqrt{\frac{29G_N}{180\pi}},
\end{equation}
with
\begin{equation}
\Lambda_{\rm eff}(\ell_{\min})
=
-\frac{270\pi}{29G_N}.
\end{equation}

Since
\[
\Lambda_{\rm eff}\to+\infty
\qquad
\text{as}
\qquad
\ell\to0,
\]
while
\[
\Lambda_{\rm eff}\to0^{-}
\qquad
\text{as}
\qquad
\ell\to\infty,
\]
this stationary point is the global minimum of the minimally coupled
AdS branch. It is clearly visible in
Fig.~\ref{fig:Lambda_eff_comparison} as the pronounced dip of the
curve.

The minimum also provides a useful quantitative measure of the
relative magnitude of the quantum correction. At
$\ell=\ell_{\min}$, one finds
\begin{equation}
\Lambda_{\rm class}(\ell_{\min})
=
-\frac{3}{\ell_{\min}^{2}},
\qquad
\Delta\Lambda(\ell_{\min})
=
\frac{3}{2\ell_{\min}^{2}},
\end{equation}
so that
\begin{equation}
\Delta\Lambda(\ell_{\min})
=
\frac{1}{2}
\left|
\Lambda_{\rm class}(\ell_{\min})
\right|.
\label{eq:quantum_half_classical}
\end{equation}
Thus, at the minimum, the quantum correction amounts to one half of
the magnitude of the classical AdS contribution. Within the present
semiclassical description, $\ell_{\min}$ therefore identifies a
characteristic scale at which the quantum correction is already
substantial: it reduces the magnitude of the net negative effective
cosmological constant by a factor of two, while the total remains
negative. This statement is purely a quantitative characterization
of the relative contributions at the stationary point and does not,
by itself, imply dynamical stability.

For the conformally coupled de Sitter field, $A=3$ and
$B=+G_N/(120\pi)$, so that $AB>0$ and Eq.~\eqref{eq:lcrit} has no real
solution. The same holds for the minimally coupled de Sitter field,
with $A=3$ and $B=+61G_N/(120\pi)$, and for the conformally coupled
Anti-de Sitter field, with $A=-3$ and
$B=-G_N/(120\pi)$. In these three cases, the classical and quantum
contributions have the same sign, and $\Lambda_{\rm eff}(\ell)$ is
monotonic.

The second characteristic scale is obtained by examining where the
function intersects the horizontal axis. This occurs when
\begin{equation}
\Lambda_{\rm eff}(\ell)=0.
\end{equation}
Using Eq.~\eqref{eq:universal_structure}, the location of this
intersection is determined by
\begin{equation}
\ell_0^2
=
-\frac{B}{A}.
\label{eq:lzero}
\end{equation}
A real intersection with the $\Lambda_{\rm eff}=0$ axis therefore
exists only when $AB<0$, the same sign condition that allows the
existence of an extremum. At this point, the two contributions to the
effective cosmological constant exactly compensate each other,
\begin{equation}
\Lambda_{\rm class}(\ell_0)
+
\Delta\Lambda(\ell_0)
=
0,
\end{equation}
or equivalently,
\begin{equation}
\Delta\Lambda(\ell_0)
=
-\Lambda_{\rm class}(\ell_0).
\end{equation}
Thus, $\ell_0$ identifies the characteristic scale at which the
quantum correction reaches precisely the magnitude required to cancel
the classical contribution.

For the minimally coupled AdS field, Eq.~\eqref{eq:lzero} reproduces
the intersection already given in Eq.~\eqref{eq:ell_zero_ads}.

Indeed, at this value,
\begin{equation}
\Lambda_{\rm class}(\ell_0)
=
-\frac{3}{\ell_0^2},
\qquad
\Delta\Lambda(\ell_0)
=
\frac{29G_N}{120\pi\ell_0^4},
\end{equation}
so that
\begin{equation}
\Delta\Lambda(\ell_0)
=
-\Lambda_{\rm class}(\ell_0),
\end{equation}
and hence
\begin{equation}
\Lambda_{\rm eff}(\ell_0)=0.
\end{equation}

The zero at $\ell_0$ reflects the exact cancellation of the classical
and quantum contributions to $\Lambda_{\rm eff}$, evaluated on the
prescribed AdS background. It does not imply that the geometry itself
becomes flat at $\ell_0$. The relation $\ell_0 = \ell_{\min}/\sqrt{2}$
fixes the relative position of the two scales in the minimally coupled
AdS curve, whose behaviour has been described in
Sec.~\ref{subsec:competition}.

For the three configurations with $AB>0$, the same search for an
intersection gives $\ell_0^2<0$, so the corresponding curves have no
real intersection with the $\Lambda_{\rm eff}=0$ axis. This is
consistent with the classification of
Sec.~\ref{subsec:classification}: the classical and quantum
contributions share the same sign and therefore cannot cancel for any
real $\ell$. Consequently, $\Lambda_{\rm eff}(\ell)$ remains strictly
positive for the dS configurations and strictly negative for the AdS
conformal one. This behaviour is visible in
Fig.~\ref{fig:Lambda_eff_comparison}.

% ============================================================
\subsection{Relative Strength of Quantum Corrections}
\label{subsec:relative_strength}
% ============================================================

The relative importance of vacuum polarization can be quantified through
the dimensionless ratio
\begin{equation}
\left|
\frac{\Delta\Lambda}{\Lambda_{\rm class}}
\right|
=
\left|
\frac{B}{A}
\right|
\frac{1}{\ell^{2}}.
\label{eq:relative_correction}
\end{equation}

Since the classical contribution scales as $\ell^{-2}$ whereas the
vacuum-polarization contribution scales as $\ell^{-4}$, their relative
magnitude scales as $\ell^{-2}$. Thus, quantum corrections become
progressively suppressed as the curvature radius increases and become
increasingly important relative to the classical contribution as
$\ell$ decreases.

To quantify when the quantum correction remains small compared with the
classical contribution, we introduce a threshold scale
$\ell_{\epsilon}$ defined by
\begin{equation}
\left|
\frac{\Delta\Lambda}{\Lambda_{\rm class}}
\right|
=
\epsilon
\quad\Longleftrightarrow\quad
\frac{|C|G_{\rm N}}{360\pi\ell^{2}}
=
\epsilon,
\label{eq:threshold_condition}
\end{equation}
where $|C|$ denotes the magnitude of the coefficient controlling the
relative quantum correction. Solving for the corresponding curvature
radius gives
\begin{equation}
\ell_{\epsilon}
=
\sqrt{\frac{|C|}{360\pi\epsilon}}\,
\ell_{\rm Pl},
\label{eq:threshold_scale}
\end{equation}
where $\ell_{\rm Pl}=\sqrt{G_{\rm N}}$. Thus,
$\ell\geq\ell_{\epsilon}$ corresponds to
$|\Delta\Lambda/\Lambda_{\rm class}|\leq\epsilon$, whereas
$\ell<\ell_{\epsilon}$ corresponds to a correction exceeding the chosen
threshold.

We use $\epsilon=0.01$ as a reference criterion for a small quantum
correction. This choice provides a convenient quantitative benchmark
for comparing the relative strength of vacuum polarization among the
different configurations; it is not intended as a fundamental boundary
of validity of semiclassical gravity.

\begin{figure}[htbp]
    \centering
    \includegraphics[width=0.85\textwidth]{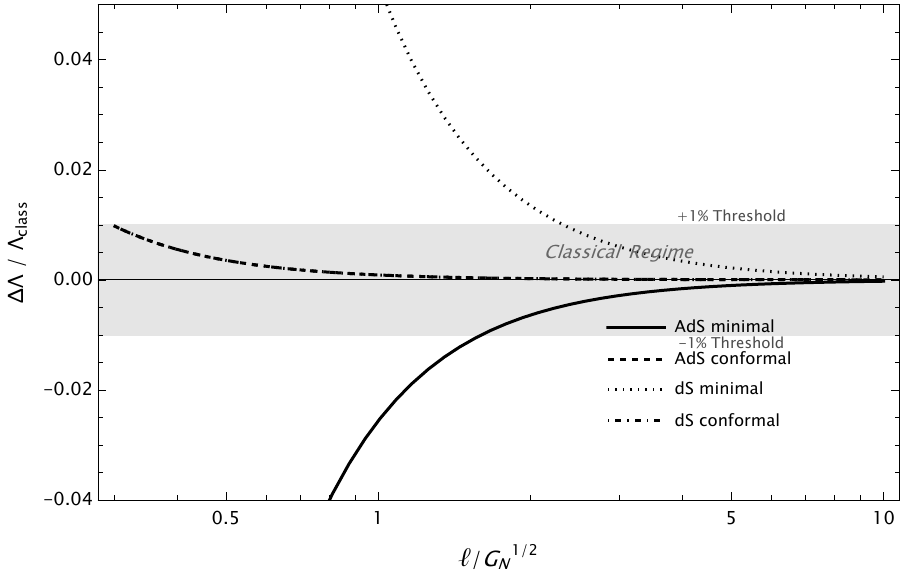}
    \caption{Relative quantum correction to the classical cosmological
    contribution, shown as
    $\Delta\Lambda/\Lambda_{\rm class}$ as a function of
    $\ell/\ell_{\rm Pl}$ on a logarithmic horizontal scale.
    The shaded band corresponds to
    $|\Delta\Lambda/\Lambda_{\rm class}|\leq0.01$, adopted here as a
    reference criterion for a small quantum correction. The AdS minimal
    case is negative, whereas the dS minimal, AdS conformal, and dS
    conformal cases yield positive relative corrections. The two
    conformal cases are exactly degenerate because
    $B/A=G_{\rm N}/(360\pi)$ for both geometries.}
    \label{fig:Lambda_threshold}
\end{figure}

Figure~\ref{fig:Lambda_threshold} provides a direct graphical
representation of the relative quantum correction and its dependence on
the curvature radius. The horizontal axis is expressed in units of the
Planck length, $\ell/\ell_{\rm Pl}$, while the vertical axis shows the
signed ratio $\Delta\Lambda/\Lambda_{\rm class}$. The shaded band
corresponds to
$|\Delta\Lambda/\Lambda_{\rm class}|\leq0.01$, thereby identifying the
region in which the quantum correction remains below the adopted $1\%$
reference criterion.

The intersections of the curves with the boundaries of this band
reproduce the threshold scales obtained analytically above. In
particular, the dS minimal configuration, which exhibits the largest
positive relative correction, crosses the upper $+1\%$ boundary at
$\ell\simeq2.32\,\ell_{\rm Pl}$. Thus, for
$\ell\lesssim2.32\,\ell_{\rm Pl}$, its relative quantum correction
exceeds the $1\%$ reference threshold. The AdS minimal configuration
behaves analogously on the negative side: its curve crosses the lower
$-1\%$ boundary at
$\ell\simeq1.60\,\ell_{\rm Pl}$, so that the magnitude of its relative
quantum correction exceeds $1\%$ for
$\ell\lesssim1.60\,\ell_{\rm Pl}$.

The two conformal configurations are exactly degenerate in the relative
correction, since both satisfy
$B/A=G_{\rm N}/(360\pi)$. Their curves therefore coincide and cross the
upper $+1\%$ boundary at
$\ell\simeq0.30\,\ell_{\rm Pl}$. Consequently, the conformal cases
remain within the $1\%$ reference band down to substantially smaller
curvature radii than either of the minimally coupled configurations.

The figure therefore makes explicit the hierarchy already encoded in
the coefficients of Table~\ref{tab:AB}: the dS minimal configuration
has the largest relative quantum correction, followed by the AdS
minimal configuration, while the two conformal cases exhibit the
smallest relative correction. This ordering is controlled by
$|C|=61$, $29$, and $1$, respectively, and follows directly from the
$\ell^{-2}$ scaling of the relative correction. The agreement between
the analytic threshold values and the corresponding intersections in
Fig.~\ref{fig:Lambda_threshold} provides a consistency check on the
quantitative characterization of the semiclassical regime. The
$1\%$ criterion should, however, be understood solely as a reference
benchmark adopted for this comparison, rather than as a fundamental
boundary separating classical and semiclassical gravity.

\subsection{Self-Consistent Semiclassical Vacua}
\label{subsec:self_consistent_vacua}

As established in Secs.~\ref{subsec:ads_selfconsistent} and
\ref{subsec:ds_selfconsistent}, both geometries admit self-consistent
configurations selected by the vanishing of the renormalized
stress--energy tensor, corresponding to $\Delta\Lambda=0$ in the
decomposition~\eqref{eq:LambdaEff_decomp}. We now compare the
resulting parameter constraints in the two spacetimes.

As derived in Sec.~\ref{sec:ds}, the condition
$\langle T_{\mu\nu}\rangle_{\rm ren}=0$ reduces in de Sitter
spacetime to $F(m^2,\xi,R)=0$ (Eq.~\eqref{eq:Fzero_ds}), yielding
for minimal coupling the self-consistent value
$m^2 = 4.89707\times10^{12}\Lambda$
(Eq.~\eqref{eq:root_ds}). Following the AdS analysis of
Sec.~\ref{sec:ads}, the same condition reduces to
$f_\xi^{(D)}(y)=0$ (Eq.~\eqref{eq:selfconsistency_ads}), with the
minimal Dirichlet solution $m^2 = -0.4014362786\,\Lambda$
(Eq.~\eqref{eq:ads_mass_lambda}).

The contrast between these two results is striking. Although both
emerge from the same semiclassical requirement
$\langle T_{\mu\nu}\rangle_{\rm ren}=0$, the mass--curvature
relations differ by more than thirteen orders of magnitude at equal
magnitude of the cosmological constant,
\[
\frac{m^2_{\rm dS}}{m^2_{\rm AdS}} \sim 10^{13}.
\]
This disparity is consistent with the distinct global and infrared
structures of the two quantum constructions. As discussed in
Sec.~\ref{sec:ds_massless}, the dS infrared sector is determined by
the Bunch--Davies state and the cosmological horizon. In AdS, as shown in Sec.~\ref{sec:ads}, it is instead controlled by
the timelike conformal boundary, and the resulting self-consistent
spectrum depends explicitly on the admissible boundary conditions.

\section{Analytic Continuation and the Limits of Curvature Correspondence}
\label{sec:analytic_cont}

The relation between Anti-de Sitter and de Sitter spacetimes through analytic continuation \cite{martin2020higher} has long been a useful formal tool in gravitational physics. At the level of classical geometry, the two maximally symmetric spaces are related by the continuation of the curvature radius,
\begin{equation}
\ell \longrightarrow i\ell .
\end{equation}
Under this transformation, the curvature scales behave as
\begin{align}
\ell^2 &\longrightarrow -\ell^2,\\
\ell^4 &\longrightarrow \ell^4 .
\end{align}
Consequently, quantities depending only on the local curvature scale may exhibit a direct correspondence between de Sitter and Anti-de Sitter geometries \cite{balasubramanian2003exploring}. However, semiclassical gravity contains additional information beyond the local curvature tensor: the renormalized stress-energy tensor depends on the global quantum state, boundary conditions, and infrared properties of the quantum field.

In this section, we investigate to what extent the analytic continuation of the curvature radius reproduces the semiclassical backreaction obtained independently in Anti-de Sitter spacetime. We shall see that, in general, the quantum correction is not preserved, and that the failure manifests itself in different ways for conformal and minimal coupling.

\subsection{Conformal Coupling: Sign Flip of the Quantum Correction}

We first consider the conformally coupled scalar field, for which the massless semiclassical corrections are given by Eqs.~\eqref{eq:LambdaEff_ds_conformal} and \eqref{eq:LambdaEff_AdS_conf}:
\begin{equation}
\Lambda_{\rm eff}^{\rm dS,conf}
=
\frac{3}{\ell^2}
+
\frac{G_{\rm N}}
{120\pi\ell^4},
\label{eq:analytic_dS_conf_original}
\end{equation}
and
\begin{equation}
\Lambda_{\rm eff}^{\rm AdS,conf}
=
-\frac{3}{\ell^2}
-
\frac{G_{\rm N}}
{120\pi\ell^4}.
\label{eq:analytic_AdS_conf_original}
\end{equation}

Applying the analytic continuation \(\ell\rightarrow i\ell\) to Eq.~\eqref{eq:analytic_dS_conf_original} yields
\begin{equation}
\Lambda_{\rm eff}^{\rm dS,conf}(\ell\rightarrow i\ell)
=
-\frac{3}{\ell^2}
+
\frac{G_{\rm N}}
{120\pi\ell^4}.
\end{equation}
This does not coincide with the direct Anti-de Sitter result \eqref{eq:analytic_AdS_conf_original}: the quantum correction changes sign. The discrepancy is quantified by the ratio of the quantum coefficients,
\begin{equation}
\frac{
B_{\rm AdS}^{\rm conf}
}{
B_{\rm dS}^{\rm conf}
}
=
\frac{-G_N/(120\pi)}{+G_N/(120\pi)}
=
-1,
\label{eq:analytic_conf_discrepancy}
\end{equation}
where the coefficients are taken from Table~\ref{tab:AB}. Thus, in the conformal sector, the analytic continuation fails to preserve the sign of the quantum correction, although the magnitude is correctly reproduced.

\subsection{Minimal Coupling: Magnitude Mismatch}

The situation is even more severe for a minimally coupled scalar field. In de Sitter spacetime, the massless correction is given by Eq.~\eqref{eq:LambdaEff_ds_minimal}:
\begin{equation}
\Lambda_{\rm eff}^{\rm dS,min}
=
\frac{3}{\ell^2}
+
\frac{61\,G_{\rm N}}
{120\pi\ell^4}.
\label{eq:analytic_dS_min_original}
\end{equation}
The same analytic continuation gives
\begin{equation}
\Lambda_{\rm eff}^{\rm dS,min}(\ell\rightarrow i\ell)
=
-\frac{3}{\ell^2}
+
\frac{61\,G_{\rm N}}
{120\pi\ell^4}.
\label{eq:analytic_dS_min_continued}
\end{equation}
However, the direct Anti-de Sitter calculation yields
\begin{equation}
\Lambda_{\rm eff}^{\rm AdS,min}
=
-\frac{3}{\ell^2}
+
\frac{29\,G_{\rm N}}
{120\pi\ell^4}.
\label{eq:analytic_AdS_min_direct}
\end{equation}

The analytic continuation reproduces the classical curvature contribution but fails to reproduce the magnitude of the quantum correction, while preserving its sign. The discrepancy is quantified by the ratio of the quantum coefficients,
\begin{equation}
\frac{
B_{\rm AdS}^{\rm min}
}{
B_{\rm dS}^{\rm min}
}
=
\frac{29}{61},
\label{eq:analytic_min_discrepancy}
\end{equation}
where again the coefficients are taken from Table~\ref{tab:AB}. This mismatch demonstrates that the semiclassical vacuum energy is not determined solely by the local curvature scale.

\subsection{Global Origin of the Breakdown}

The failure of the analytic continuation in both the conformal and minimal sectors is related to the global information entering the definition of the quantum state.

As already discussed in Sec.~\ref{sec:ds_massless}, the dS infrared
sector is determined by the Bunch--Davies state and by the global
causal structure of the spacetime, including the presence of a
cosmological horizon. These distinct infrared behaviours are encoded
in the different numerical coefficients of Table~\ref{tab:AB}.

In AdS, the infrared sector is instead controlled by the timelike
conformal boundary. Following the analysis of Sec.~\ref{sec:ads},
the conformal and minimal sectors exhibit distinct behaviours in the
massless limit.

Therefore, the analytic transformation $\ell\rightarrow i\ell$ correctly maps the local curvature scale,
\begin{equation}
R_{\rm dS}
=
\frac{12}{\ell^2}
\longrightarrow
R_{\rm AdS}
=
-\frac{12}{\ell^2},
\end{equation}
but it does not automatically map the complete quantum theory. In particular, it does not preserve the information associated with the choice of quantum state and boundary conditions. The factors \(-1\) and \(29/61\) in Eqs.~\eqref{eq:analytic_conf_discrepancy} and \eqref{eq:analytic_min_discrepancy} are quantitative measures of this failure.

Hence, analytic continuation provides a reliable correspondence only at the level of classical geometry. It does not in general preserve the full vacuum structure of the theory. Both the conformally and minimally coupled scalar fields provide explicit examples in which two maximally symmetric spacetimes related by curvature continuation exhibit substantially different semiclassical backreaction, demonstrating that global quantum properties cannot be reconstructed from local curvature data alone.

\section{Conclusions}
\label{sec:conclusions}

We have investigated the semiclassical Einstein equation for a
quantized scalar field in four-dimensional anti-de Sitter (AdS) and
de Sitter (dS) spacetimes from two complementary perspectives. First,
we imposed self-consistency while fixing the bare cosmological
constant to its geometric value,
$\Lambda_{\rm B}=\Lambda_{\rm geom}$. The geometric and bare
cosmological-constant terms then cancel in the semiclassical
Einstein equation, requiring the renormalized vacuum
stress--energy tensor to vanish. This condition does not imply the
absence of quantum effects; rather, it selects field parameters for
which the quantum vacuum is compatible with the prescribed geometry
without shifting the bare cosmological constant. Second, we evaluated
the vacuum contribution to the effective cosmological constant on
prescribed maximally symmetric backgrounds. These two analyses
distinguish self-consistent parameter constraints from the quantum
corrections associated with a fixed background.

The self-consistent solutions depend on the spacetime and the
quantum construction. In dS, for minimal coupling, we recover the
previously reported solution for a massive scalar field. In AdS, the
solutions depend additionally on the admissible boundary conditions.
For minimal coupling, the Dirichlet condition yields a self-consistent
solution, whereas the formal Neumann root lies outside the allowed
parameter range. For conformal coupling, both Dirichlet and Neumann
solutions are admissible. In particular, the Neumann solution has
negative mass squared while satisfying the Breitenlohner--Freedman
bound. Thus, a negative mass-squared parameter in AdS does not by
itself imply an instability; its admissibility must be assessed
against the relevant stability bound and boundary conditions.

For the massless limits considered, the effective cosmological
constant takes the common form
\[
\Lambda_{\rm eff}(\ell)
=\frac{A}{\ell^2}+\frac{B}{\ell^4},
\]
where the first term represents the classical geometric
contribution and the second the renormalized vacuum contribution in
the convention adopted here. The signs of these terms determine the
qualitative behaviour of $\Lambda_{\rm eff}$. In minimally coupled
AdS, the two contributions have opposite signs, producing a minimum
and a zero at finite curvature radii. In the other three cases
considered, the terms have the same sign and the effective quantity
is monotonic. These extrema and zeros characterize
$\Lambda_{\rm eff}$ evaluated on the prescribed backgrounds; they
do not, by themselves, establish a dynamical transition between
different spacetime geometries.

The coefficients of the quantum contributions also reveal a
quantitative mismatch between the AdS and dS results. For minimal
coupling,
\[
B_{\rm dS}^{\rm min}=\frac{61G_N}{120\pi},
\qquad
B_{\rm AdS}^{\rm min}=\frac{29G_N}{120\pi},
\qquad
\frac{B_{\rm AdS}^{\rm min}}{B_{\rm dS}^{\rm min}}
=\frac{29}{61}.
\]
For conformal coupling,
\[
B_{\rm dS}^{\rm conf}=\frac{G_N}{120\pi},
\qquad
B_{\rm AdS}^{\rm conf}=-\frac{G_N}{120\pi}.
\]
Consequently, the minimal-coupling coefficients differ in magnitude,
whereas the conformal-coupling coefficients have equal magnitudes
and opposite signs. The formal continuation
$\ell\rightarrow i\ell$ reproduces the sign change of the classical
curvature term, but does not reproduce these directly calculated
quantum coefficients. Within the constructions and renormalization
prescription considered here, continuation of the curvature scale
alone is therefore insufficient to establish a correspondence
between the full semiclassical vacuum contributions in AdS and dS.

The relative magnitude of the quantum correction scales as
$\ell^{-2}$ compared with the classical contribution. Using a
$1\%$ relative correction as a reference criterion gives characteristic
radii of approximately $2.32\,\ell_{\rm Pl}$ for minimal coupling in
dS, $1.60\,\ell_{\rm Pl}$ for minimal coupling in AdS, and
$0.30\,\ell_{\rm Pl}$ for both conformal cases. These values provide
a comparative measure of the relative importance of vacuum
polarization, rather than fundamental boundaries of validity.
In particular, the sub-Planckian conformal scales should not be
interpreted as evidence that the semiclassical approximation remains
quantitatively reliable at those scales.

The comparison must be understood in light of the distinct quantum
structures of the two spacetimes. In AdS, the timelike conformal
boundary makes the choice of boundary condition part of the field
theory's definition. In dS, the choice of quantum state and the
infrared behaviour of the field play a central role. In particular,
the massless minimally coupled dS result obtained here is the
one-sided limit $m\to0^+$ of the massive Bunch--Davies expression;
it is not the expectation value in a fully de Sitter-invariant Fock
vacuum for the exactly massless, minimally coupled field. These
qualifications are essential when interpreting the comparison and
the analytic continuation between the two backgrounds.

Several extensions would be worthwhile. One is to investigate
analytic continuation at the level of the underlying two-point
functions, rather than only through the renormalized
stress--energy tensor. This could provide a more direct test of
whether a relation between the AdS and dS quantum constructions
can be established. Further directions include exploring other
admissible AdS boundary conditions and quantum states, extending
the analysis to other quantum fields and spacetime dimensions, and
determining how interactions or additional fields modify the
effective cosmological contribution and the mismatch between the
AdS and dS coefficients.

In summary, the self-consistent analysis identifies field parameters
compatible with an unchanged maximally symmetric geometry at fixed
bare cosmological constant, while the effective-cosmological-constant
analysis quantifies the vacuum contribution on prescribed
backgrounds. Their comparison establishes both a common
curvature-dependent structure and a quantitative failure of the
formal continuation $\ell\rightarrow i\ell$ to reproduce the
calculated quantum coefficients. The results underscore the
importance of specifying the quantum state, admissible boundary
conditions, and global structure when evaluating semiclassical
vacuum contributions to the cosmological constant.

\section*{Acknowledgments}

We are grateful to Benito A. Juárez-Aubry for illuminating discussions and continuous encouragement. M.C.M.-L. also thanks his wife Thayss for her unconditional support.

% ====== BIBLIOGRAPHY ======
\bibliographystyle{unsrt}
\bibliography{paper}

@article{Weinberg1989,
  author = {S. Weinberg},
  title = {The Cosmological Constant Problem},
  journal = {Rev. Mod. Phys.},
  volume = {61},
  pages = {1},
  year = {1989},
  doi = {10.1103/RevModPhys.61.1}
}

@article{Padmanabhan2003,
  author = {T. Padmanabhan},
  title = {Cosmological Constant: The Weight of the Vacuum},
  journal = {Phys. Rept.},
  volume = {380},
  pages = {235},
  year = {2003},
  doi = {10.1016/S0370-1573(03)00120-0},
  eprint = {hep-th/0212290}
}

@article{Martin2012,
  author = {J. Martin},
  title = {Everything You Always Wanted to Know About the Cosmological Constant Problem},
  journal = {Comptes Rendus Physique},
  volume = {13},
  pages = {566},
  year = {2012},
  doi = {10.1016/j.crhy.2012.04.008},
  eprint = {1205.3365}
}

@book{BirrellDavies,
  author = {N. D. Birrell and P. C. W. Davies},
  title = {Quantum Fields in Curved Space},
  publisher = {Cambridge University Press},
  year = {1982}
}

@book{WaldQFT,
  author = {R. M. Wald},
  title = {Quantum Field Theory in Curved Spacetime and Black Hole Thermodynamics},
  publisher = {University of Chicago Press},
  year = {1994}
}

@article{Allen1985,
  author = {B. Allen},
  title = {Vacuum States in de Sitter Space},
  journal = {Phys. Rev. D},
  volume = {32},
  pages = {3136},
  year = {1985},
  doi = {10.1103/PhysRevD.32.3136}
}

@article{JuarezAubry2019,
  author = {B. A. Ju\'arez-Aubry},
  title = {Semi-classical Gravity in de Sitter Spacetime and the Cosmological Constant},
  journal = {Phys. Lett. B},
  volume = {797},
  pages = {134912},
  year = {2019},
  doi = {10.1016/j.physletb.2019.134912},
  eprint = {1903.03924}
}

@article{JuarezAubryMamani2025,
  author = {B. A. Ju\'arez-Aubry and M. C. Mamani-Leqque},
  title = {Renormalisation in Maximally Symmetric Spaces and Semiclassical Gravity in Anti-de Sitter Spacetime},
  journal = {Class. Quant. Grav.},
  volume = {42},
  pages = {085009},
  year = {2025},
  doi = {10.1088/1361-6382/adc535},
  eprint = {2412.00000}
}

@article{gottschalk2022cosmological,
  title={Cosmological de Sitter solutions of the semiclassical Einstein equation},
  author={Gottschalk, Hanno and Rothe, Nicolai and Siemssen, Daniel},
  journal={Ann. Henri Poincaré},
  volume={24},
  pages={2949},
  year={2023},
  doi={10.1007/s00023-023-01315-z}
}

@article{hollands2001local,
  title={Local Wick Polynomials and Time Ordered Products  of Quantum Fields in Curved Spacetime},
  author={Hollands, Stefan and Wald, Robert M},
  journal={Communications in Mathematical Physics},
  volume={223},
  number={2},
  pages={289--326},
  year={2001},
  publisher={Springer}
}

@article{breitenlohner1982stability,
  title={Stability in gauged extended supergravity},
  author={Breitenlohner, Peter and Freedman, Daniel Z},
  journal={Annals of physics},
  volume={144},
  number={2},
  pages={249--281},
  year={1982},
  publisher={Elsevier}
}

@article{dappiaggi2016hadamard,
  title={Hadamard states for a scalar field in anti--de Sitter spacetime with arbitrary boundary conditions},
  author={Dappiaggi, Claudio and Ferreira, Hugo RC},
  journal={Physical Review D},
  volume={94},
  number={12},
  pages={125016},
  year={2016},
  publisher={APS}
}

@article{decanini2008hadamard,
  title={Hadamard renormalization of the stress-energy tensor for a quantized scalar field in a general spacetime of arbitrary dimension},
  author={D{\'e}canini, Yves and Folacci, Antoine},
  journal={Physical Review D},
  volume={78},
  number={4},
  pages={044025},
  year={2008},
  publisher={APS}
}

@article{martin2020higher,
  title={Higher spin partition functions via the quasinormal mode method in de Sitter quantum gravity},
  author={Martin, Victoria and Svesko, Andrew},
  journal={SciPost Physics},
  volume={9},
  number={3},
  pages={039},
  year={2020}
}

@article{balasubramanian2003exploring,
  title={Exploring de Sitter space and holography},
  author={Balasubramanian, Vijay and de Boer, Jan and Minic, Djordje},
  journal={Annals of Physics},
  volume={303},
  number={1},
  pages={59--116},
  year={2003},
  publisher={Elsevier}
}

@article{pitelli2019boundary,
  title={Boundary conditions and renormalized stress-energy tensor on a Poincar{\'e} patch of AdS 2},
  author={Pitelli, Jo{\~a}o Paulo M and Barroso, Vitor S and Mosna, Ricardo A},
  journal={Physical Review D},
  volume={99},
  number={12},
  pages={125008},
  year={2019},
  publisher={APS}
}

@article{ford2002does,
  title={What does quantum field theory in curved spacetime have to say about the dark energy?},
  author={Ford, LH},
  journal={arXiv preprint gr-qc/0210096},
  year={2002}
}

@article{bunch1978quantum,
  title={Quantum field theory in de Sitter space: renormalization by point-splitting},
  author={Bunch, Timothy S and Davies, Paul CW},
  journal={Proceedings of the Royal Society of London. Series A, Mathematical and Physical Sciences},
  pages={117--134},
  year={1978},
  publisher={JSTOR}
}

@article{strominger2001ds,
  title={The dS/CFT correspondence},
  author={Strominger, Andrew},
  journal={Journal of High Energy Physics},
  volume={2001},
  number={10},
  pages={034--034},
  year={2001}
}

\end{document}